\documentclass[%
 aip,
rsi,
 amsmath,amssymb,
 reprint,%
]{revtex4-1}

\usepackage{graphicx}
\graphicspath{{Images/}}
\usepackage{dcolumn}
\usepackage{bm}

\usepackage[utf8]{inputenc}
\usepackage[T1]{fontenc}
\usepackage{mathptmx}
\usepackage{etoolbox}
\usepackage{enumitem}

\makeatletter
\def\@email#1#2{%
 \endgroup
 \patchcmd{\titleblock@produce}
  {\frontmatter@RRAPformat}
  {\frontmatter@RRAPformat{\produce@RRAP{*#1\href{mailto:#2}{#2}}}\frontmatter@RRAPformat}
  {}{}
}%
\makeatother
\begin{document}

\preprint{AIP/123-QED}

\title[]{Portable Real-Time Polarimeter for Partially and Fully Polarized Light}
\author{B. Mackey}
 \affiliation{%
 Department of Physics and Astronomy, University of Victoria, British Columbia, Canada
 }%
\author{O. Sandner}
 \affiliation{%
 Department of Physics and Astronomy, University of Victoria, British Columbia, Canada
 }%
\author{A. Saji}
 \affiliation{%
 International School of Photonics, Cochin University of Science and Technology, India
 }%
\author{A. Felipe Ramos}
 \affiliation{%
 Universidad del Valle, Santiago de Cali, Valle del Cauca, Colombia
 }%
 \author{L. Hall}
 \affiliation{%
 Department of Physics and Astronomy, University of Victoria, British Columbia, Canada
 }%
\author{S. Wilkinson}
 \affiliation{%
 Department of Physics and Astronomy, University of Victoria, British Columbia, Canada
 }%
\author{A. Mackay}
 \affiliation{%
 Department of Physics and Astronomy, University of Victoria, British Columbia, Canada
 }%
\author{Nicolas Braam}
\affiliation
{%
 Department of Physics and Astronomy, University of Victoria, British Columbia, Canada
}%
\author{Chris Secord} 
\affiliation{%
 Faculty of Science Machine Shop, University of Victoria, British Columbia, Canada
 }%
\author{A. MacRae}
 \email{macrae@uvic.ca}
\affiliation{%
Department of Physics and Astronomy, University of Victoria, British Columbia, Canada
}%

\date{\today}

\begin{abstract}
We present a portable polarimeter capable of real-time visualization of partial and fully polarized light over a broad band of wavelengths. Our system utilizes a Raspberry Pi computer with a low-cost data acquisition "HAT" (DAQ HAT) and an integrated photodetection circuit. Wide bandwidth operation is achieved through digital calibration of an arbitrary retardance waveplate presented herein. All mechanical, electrical, and software components are open source and available on a GitHub repository. This completely integrated approach provides an efficient tool for modern optics research laboratories and is well-suited for educational demonstrations.
\end{abstract}

\maketitle

\section{\label{sec:intro} Introduction}

The quantification of optical polarization plays a key role in the analysis of numerous physical processes, with applications in Atomic, Molecular, and Optical (AMO) Physics~\cite{garcia2018improving}, Astronomy \cite{marshall2016polarization}, Imaging \cite{losurdo2009spectroscopic}, and Material Science \cite{capelle2017polarimetric}. Monitoring the polarization of light produced in a process can reveal insights into its underlying mechanisms. Furthermore, multiple optical technologies, such as polarization maintaining fibers, optical trapping, and photonic waveguides, require precise tuning of the input polarization state. Thus, having a portable, tunable, real-time polarimeter is useful in many educational and research applications.

The polarization of an optical state may be inferred by multiple measurements of the time averaged intensity using standard polarizing material (see section~\ref{subsec:stokes}).
Such measurements generally fall into two categories: spatial modulation and time-dependent. In spatial modulation techniques, the beam is sent to a spatially varying polarizing element and detected on a CCD array. Previously, micro-mirror arrays have been used to split light into four beams corresponding to the Stokes measurements \cite{zhao_19}, and conical mirrors have been used to create a continuous signal based upon total internal reflection \cite{hawley_19}.

In time dependent measurements, a polarizing element is varied with time as the intensity is recorded. These methods typically involve a linear polarizer and a modulated birefringent element, such as a liquid crystal variable retarder \cite{bueno_00} or a rotating quarter waveplate (QWP) sampled at discrete points \cite{berry_77,schaefer2007measuring} or continuously \cite{bobach2017note,wilkinson2021complete}. This latter approach is often referred to as a spinning waveplate polarimeter (SWP) and is the method used in this work.

This paper presents an affordable, integrated SWP capable of real-time acquisition and visualization of the polarization state of an optical field. The system is easily calibrated to accommodate a broad range of intensities and wavelengths.

Our device has advantages over similar designs\cite{bobach2017note, wilkinson2021complete, chaturvedi2022compact}; it operates over an extremely wide bandwidth (over $300$~nm in this work), is very low cost, has a customizable heads-up display, is completely open source, and has efficient calibration processes. The measured accuracy of the device is comparable with commercially available alternatives \cite{Thorlabs_polarimeter, Medowlark_Optics_Polarimeter}. Detailed designs and procedures are provided in a GitHub repository~\cite{SWP_Github}.

\section{\label{sec:quan}Experimentally Quantifying Polarization}
\subsection{\label{subsec:stokes}The Stokes Parameters}
The polarization state of a quasi-monochromatic plane wave traveling in the $\hat{z}$ direction can be completely specified by the magnitudes and relative phase of its transverse components $E_x(t)$ and $E_y(t)$. Letting the unknown relative phase between the transverse components be $\phi = \phi_y - \phi_x$, the parametric plot at a fixed location $z$ is given by:
\begin{equation}
\label{eq:monochrome}
\vec{E}(\vec{r},t) = \left(E_x\hat{x} + E_ye^{-i\phi}\hat{y}\right)e^{i(kz-\omega t - \phi_x)}.
\end{equation}
This equation generally describes \textit{elliptical polarization}~\cite{born2013principles}. In the special cases that the two fields are perfectly in phase, where $\phi = n\pi$, or in quadrature, where $\phi = (2n+1)\pi/2$ for integer $n$, the ellipse becomes degenerate (\textit{linear polarization}), or has zero eccentricity (\textit{circular polarization}) respectively. 

Any pure polarization state may be uniquely expressed as a superposition of two distinct polarization basis states. While there are an infinite number of such basis choices, the most common, and relevant to this work, are: horizontal and vertical linear polarization, $+45^\circ$ and $-45^\circ$ linear polarization, or right and left circular polarization.

A challenge that arises in measuring the polarization of visible light is that the ellipse evolves at frequencies of hundreds of terahertz, and so can not be imaged directly. However, it can be completely specified using standard polarizers and intensity measurements by the \textit{Stokes parameters}.  

Denoting $I_{H}$, $I_{V}$, $I_{+}$, $I_{-}$, $I_{R}$, and $I_{L}$ as the intensity of the horizontal, vertical, $+45^\circ$, $-45^\circ$, right circular, and left circular polarizations respectively, the Stokes parameters are~\cite{born2013principles}:
\begin{subequations}
\begin{align}
S_0 &= \langle E_xE_x^*\rangle +\langle E_yE_y^*\rangle &= I_H + I_V \label{eq:defstokesa}\\
S_1 &= \langle E_xE_x^*\rangle - \langle E_yE_y^*\rangle &= I_H - I_V \label{eq:defstokesb}\\
S_2 &= \langle E_xE_y^*\rangle + \langle E_yE_x^*\rangle  &= I_+ - I_- \label{eq:defstokesc}\\
S_3 &= i(\langle E_xE_y^*\rangle - \langle E_yE_x^*\rangle) &= I_R - I_L \label{eq:defstokesd}
\end{align}
\end{subequations}

\noindent where the angled brackets denote a time average over a short measurement time and we have omitted a factor of $\varepsilon_0c$ for the middle terms.

These parameters can be encompassed in a single vector known as the \textit{Stokes vector}:
\begin{equation}
\label{eq:def_stokes_vec}
\vec{S} \equiv \left(S_0,S_1,S_2,S_3\right).
\end{equation}
The Stokes vector is useful in mathematically representing polarizing elements using the Mueller formalism~\cite{born2013principles,ware2015physics}, in which polarizing optical components are represented by a $4\times4$ Mueller matrix. Additionally, the Stokes formalism allows the quantification of partially polarized light \cite{born2013principles,ware2015physics}.  If the relative phases and amplitudes in equation~\ref{eq:monochrome} are not constant over the averaging process, $S_1$, $S_2$ and $S_3$ will average out so that in the limit of rapid variation the Stokes parameters become $S_{unp} = (1,0,0,0)$. The degree of polarization (DOP) can be written in terms of the Stokes vector as:
\begin{equation}
\label{eq:def_dop}
P \equiv \frac{\sqrt{S_1^2 + S_2^2 + S_3^2}}{S_0}.
\end{equation}
Light that is partially polarized can be written as:
\begin{equation}
\label{eq:pol_unp}
\vec{S} = P\vec{S}_{\mathrm{pol}} + (1-P)\vec{S}_{\mathrm{unp}}.
\end{equation}

\vspace{-20pt}

\subsection{\label{subsec:swp}Measuring Polarization}

\vspace{-10pt}


For a rotating polarimeter, the system matrix can be directly calculated by the Mueller formalism:
\begin{equation}
\label{eq:system_matrix}
\vec{S}^\prime = P_0R(\theta)W_\delta R(-\theta)\vec{S}.
\end{equation}
\noindent Here, $R(\theta)$ corresponds to rotation of the coordinate frame by $\theta$, $W_\delta$ is a waveplate with phase delay $\delta$ and a horizontal fast axis, and $P_0$ is a horizontally aligned linear polarizer. The output intensity is given by the first component of the Stokes vector $I(\theta) = S_0^\prime$ and is given by:
\begin{align}
\label{eq:swp_intensity}
I(\theta) &= \frac{1}{2}\left[S_0 + \left(\frac{1+\cos\delta}{2}\right)S_1\right] - \left(\frac{\sin\delta}{2}S_3\right)\sin2\theta \nonumber \\
& + \left(\frac{1-\cos\delta}{4}S_1\right)\cos4\theta + \left(\frac{1-\cos\delta}{4}S_2\right)\sin4\theta.
\end{align}

Previous works have considered a precisely calibrated QWP, but this can be difficult to achieve in practice and limits the applicability. Instead, we assume a waveplate with arbitrary retardance $\delta\ne n\pi$, $n\in\mathbb{Z}$, rotated to angle $\theta$ with respect to the horizontally aligned polarizer. Note that in the case of an ideal QWP, $\delta = \pi/2$, the expression simplifies to the expression described in previous works \cite{berry_77}:
\begin{equation}
\label{eq:swp_intensity_basic}
I(\theta) = \frac{2S_0 + S_1}{4} - \frac{S_3}{2}\sin2\theta + \frac{S_1}{4}\cos4\theta + \frac{S_2}{4}\sin4\theta.
\end{equation}

Thus, with knowledge of $\delta$ and $I(\theta)$, the Stokes parameters may be obtained algebraically with four discrete measurements. A more robust method, however, is to acquire data continuously while spinning the polarimeter at a constant rate $\theta(t) = \omega t$, and extract the Fourier components at frequencies of 0, $2\omega$, and $4\omega$.

\vspace{-10pt}
\subsection{\label{subsec:vis} Visualizing Polarization}

The polarization state of light can be represented multiples ways, each with its own advantages. The three main approaches are: \textit{a)} to make a parametric plot of the \textit{polarization ellipse}, \textit{b)} to give values or a bar-graph of the Stokes parameters, and \textit{c)} to represent the stokes parameters as points of a sphere, known as the \textit{Poincar\'e sphere}. Each of these representations are displayed in our device.

\textbf{Display of the Polarization Ellipse} can be accomplished by inferring the parameters of the ellipse with major/minor axes $a$ and $b$ respectively. Specifically these are the angle $\psi$ of $a$ with respect to the $x$ axis, and the ellipticity angle $\chi$, defined as $\tan\chi = \pm b/a$. In terms of the Stokes vectors these are given by \cite{born2013principles}:
\begin{equation}
\label{eq:defellipse}
\tan2\psi = \frac{S_2}{S_1}\hspace{15pt} \text{ and }\hspace{15pt}\sin2\chi = \frac{S_3}{S_0}.
\end{equation}

A drawback of displaying the polarization ellipse alone is that only the polarized component of the light evolves in an ellipse, so we may only visualize partial polarization by scaling down the ellipse such that $a=P$.

\textbf{Direct Display of Stokes Parameters} is most useful for partially polarized light. In this case, the parameters are either numerically displayed or drawn as a bar graph and the DOP is easily inferred. The handedness of the polarization state is directly apparent from the sign of $S_3$. The sole disadvantage of this technique is that the amount of ellipticity and angle of rotation are not readily apparent. 

\begin{figure*}
\includegraphics[width = \textwidth]{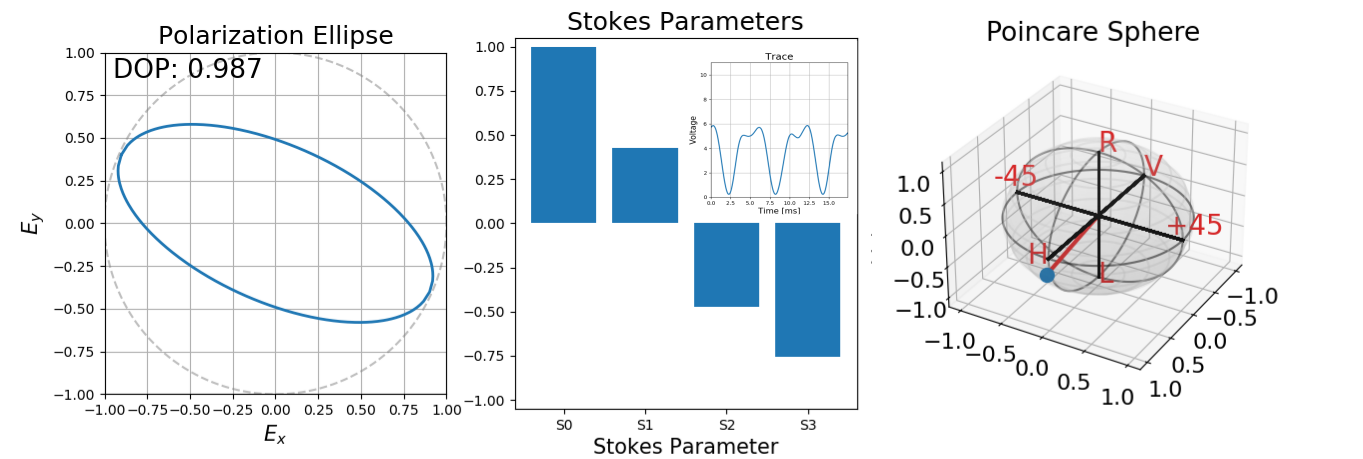}
\caption{Screen capture of the real-time display of the polarimeter as seen during operation. From left to right, the polarization ellipse is shown with the gray dashed line representing the bounding region for all polarization states. The center panel shows the corresponding Stokes parameters and the right panel displays the state on the Poincar\'{e} sphere. Alternatively, a trace of the unprocessed voltage output can replace the Poincar\'{e} sphere. Such a trace is shown in the central inset.}
\label{fig:realtime_disp}
\end{figure*}

\textbf{The Poincar\'{e} sphere} is a three dimensional representation of the polarization where each of three Cartesian axes relates to the ratio of stokes parameters $S_1$/$S_0$, $S_2$/$S_0$, or $S_3$/$S_0$.  In this formalism, the azimuthal angle $\phi_P$ and polar angle $\theta_P$, for a vector in spherical coordinates are related to the Stokes parameters by:
\begin{equation}
\label{eq:defpoincare}
\tan\phi_P = \frac{S_2}{S_1}\hspace{15pt} \text{~and~} \hspace{15pt}\cos\theta_P = \frac{S_3}{S_0}.
\end{equation}
Partial polarization my be represented by letting the length of the vector equal $P$, where fully polarized light is a vector to a point on the unit sphere ($P=1$) and unpolarized light is a point at the origin. The Poincar\'{e} sphere clearly differentiates handedness of the circular component, and is useful for mapping the evolution of a time dependant polarization. However, it can be challenging to read a projection of a three dimensional sphere on a two dimensional screen, and is therefore less useful for applications requiring real-time user interpretation.

It should be noted that each method described above is independently capable of completely specifying the polarization when combined with appropriate scaling.
Figure~\ref{fig:realtime_disp} shows the aforementioned visualizations for a given polarization state using our polarimeter.

\vspace{-10pt}
\section{\label{sec:our_pol}Our Polarimeter}
\subsection{\label{subsec:mech}Physical Construction}
\begin{figure}[ht] 
\begin{center}
\includegraphics[width = \columnwidth]{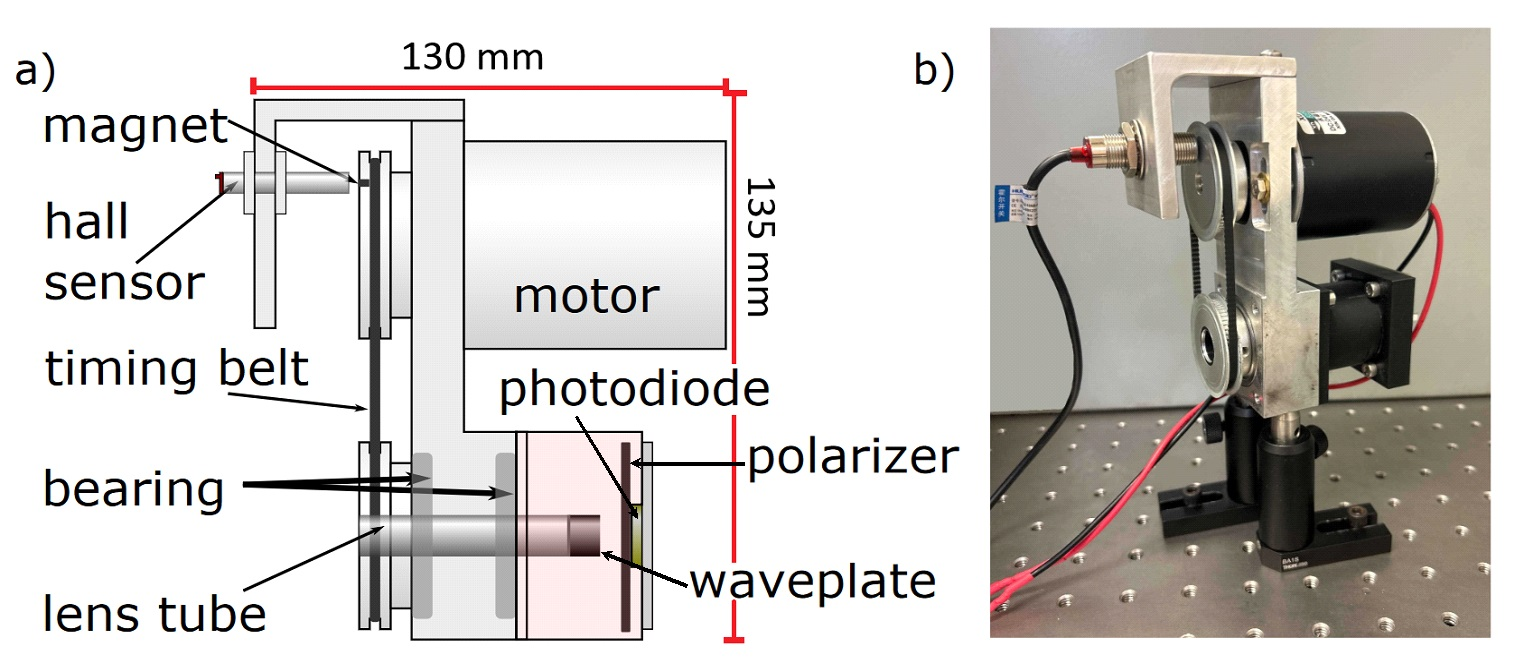}
\caption{a) Mechanical design of our SWP. b) Photograph of device on optics table.}
\label{fig:ourswp} 
\end{center}
\end{figure}

Our SWP (Figure~\ref{fig:ourswp}), consists of a XD-3420 24V brush motor connected to an externally threaded rotating tube (Thorlabs SM05M20) by a timing belt. The tube is mounted to bearings that are press-fit into the main housing. The housing contains a 0.5 inch diameter, polymer, zero-order QWP (Thorlabs WPQ05ME-780) that threads onto the tube. A small magnet is mounted near the edge of the upper timing gear that triggers a Hall sensor mounted to the main base. This magnet-sensor system provides the zero-point reference for rotation. The angle of the magnet with respect to the fast axis of the QWP is calibrated via a simple delay (see section~\ref{subsec:waveplatealignment}). The motor is controlled by a standard pulse-width modulation controller and rotates at 4500-6000 rpm. A 3D-printed adaptor plate containing a polarizing sheet is fastened to the polarimeter body and mounts the photodetection system. The photodetection system utilizes a customized circuitry system with a seven-level gain input. A detailed description of the construction is given in the supplementary materials (see section~\ref{sec:supplemental}).  


The wide availability of high quality brush motors and Raspberry Pi microcontrollers, combined with minimal optical components makes our design very affordable. This cost could be reduced by using transparency paper in the place of the QWP (see section~\ref{subsec:cheapWPs}). By integrating a small LCD screen and wireless keyboard, the system can operate completely independently of a computer.


\vspace{-25pt}
\subsection{\label{subsec:acquisition}Data Acquisition and Processing}
For data processing, the polarimeter uses the API provided by the DAQ HAT to acquire a fixed number of samples. Typically, we acquire 1000 intensity samples and 1000 Hall sensor trigger samples at $20,000$ samples per second, with a waveplate rotation rate between $75~$ to $110~$Hz. The settings are chosen to ensure that multiple rotations are captured for analysis.

To process the raw data from the DAQ hat, the trace is divided into $N$ segments,  each segment corresponding to a full rotation of the QWP. 
We assume a constant rate on the scale of one rotation so that this corresponds to equally spaced samples between $0$ and $2\pi$ radians (see the supplementary material). 
Each segment is then numerically integrated to extract the Fourier coefficients, and  Equation~\ref{eq:swp_intensity} is used to determine the corresponding Stokes parameters. 
The results of several segments are averaged to reduce statistical errors. Using the above settings, the real-time display is refreshed at a rate of approximately 4~Hz.

\subsection{\label{subsec:calib}Calibration and Experimental Imperfections}

\subsubsection{Waveplate Alignment Offset Calibration}
\label{subsec:waveplatealignment}
The angle $\theta$ in equation~\ref{eq:swp_intensity} refers to the physical angle between the transmission axis of the analysis polarizer and the fast axis of the waveplate. We must therefore identify the phase offset $\phi_0$ between the Hall sensor magnet and waveplate fast axis. To accomplish this, we note that equation ~\ref{eq:swp_intensity} predicts no $\cos2\theta$ term in the output intensity. A phase offset $\theta \rightarrow \theta + \phi_0$ will transform the sine term as:
\begin{equation}
    \sin(2(\theta + \phi_0)) = \cos2\phi_0\sin2\theta + \sin2\phi_0\cos2\theta.
\end{equation}

\noindent A nonzero cosine quadrature will thus appear unless $\phi_0 = 0$ or $\phi_0 = \pi$. Sending in partially circularly polarized light, the script that acquires a single rotation, calculates the $\cos(2(\theta+\phi^\prime))$ Fourier coefficients for all possible phases $\phi^\prime$, and records the phase offset which minimizes this coefficient.

\subsubsection{Imperfect Waveplates}
A waveplate functions by creating a relative phase shift $e^{i\delta}$ known as the retardance, between its fast and slow axes. Ideally a quarter waveplate has $\delta = \pi/2$, but its dispersive and birefringent nature leads to a wavelength dependent retardance $\delta(\lambda)$. Referencing equation~\ref{eq:swp_intensity}, we see that this can be corrected as long as $\delta$ is known for the wavelength under consideration.

Our polarimeter allows for \textit{in situ} measurement of the retardance:  By sending in light which is linearly polarized in the direction that maximizes the photodetector signal, i.e. along the SWP polarization axis, we expect a photodiode signal given by:
\begin{equation}
\label{eq:calpol_v_t}
V(\theta) = \left(\cos^4\theta + \sin^4\theta + 2\cos^2\theta\sin^2\theta\cos\delta\right)V_0
\end{equation}

\noindent where $V_0$ is the voltage in the absence of the polarizer. Differentiating, the condition for extrema are $\left[\sin4\theta\left(\cos\delta -1\right)\right] = 0$, so that $\theta = \frac{n\pi}{4}$. Noting that even $n$ yield minima and odd $n$ give maxima, equation~\ref{eq:calpol_v_t} yields:
\begin{equation}
\label{eq:calpol_final}
\delta = \arccos\left(2\frac{V_\text{min}}{V_\text{max}}-1\right).
\end{equation}

This calibration is completed using an automated script. There is notably little day-to-day variation for a particular wavelength, but large variation in retardance between wavelengths. Our polarimeter shows a phase of  $\delta \approx \pi/2$ or $\lambda/4$ at $\lambda = 780~$nm, but $\delta \approx 1.91$ at $\lambda = 635~$nm corresponding to a $\lambda/3.29$-plate.

\subsubsection{Technical Noise}
Owing to the averaging process in our Fourier analysis, random errors tend not to affect the accuracy of our system. However, non-random errors, such as a constant DC offset, \textit{will} systematically bias the determination of the polarization state. 
This is evident from equation~\ref{eq:swp_intensity} where an additional constant leads to an overestimation of the sum of $S_0$ and $S_1$, generally underestimating the DOP. 
Such DC bias may originate from unpolarized background light or dark current in the photodetector. These effects may be accounted for by directly subtracting the constant DC signal present in the absence of the light source under test. This is, again, automated with a Python script, allowing for simple recalibration when environmental conditions change.

Another source of experimental error which has plagued previous designs \cite{arnoldt2011rotating} is non-constant rotation of the QWP. In extracting the sine and cosine quadratures, the program integrates along the polarimeter angle. As such, this angle must be known at all points along the trace. This may be inferred by linear interpolation between triggers. If the angular velocity is not constant, however, there will be a residual error in this estimation. Since our SWP rotates at up to $6000$~rpm, the system is intrinsically inertially stable and significant changes to the angular speed will be gradual, over several periods. This is verified by observing the residual error from the fit of the measured voltage to the theoretical curve using a linearly polarized state showing no measurable deviations from zero. This measurement is presented in detail in the supplementary material (see section~\ref{sec:supplemental}).

\section{\label{sec:applications} Performance and Applications}
\subsection{Determining the Polarization State for Various Inputs}
To test the functionality of the SWP system, several known polarization states were evaluated and the resulting measurements compared with expectation. Unpolarized, linearly polarized, circularly polarized, and partially polarized light have been quantified to demonstrate the capabilities of the device. Various wavelengths in the visible and near IR spectrum were used to validate the polarimeter for use with a broad range of wavelengths. 

Unpolarized light was generated using a light emitting diode with $\lambda = 663 \pm 9$~nm. As expected, the polarimeter determined the state to be unpolarized, showing a DOP of $2.1\%$. The residual polarization could be due to a slight polarization by reflection at the surface of the lens used to collimate the light.
 
Linearly polarized light was generated by placing a polarizer aligned horizontally in front of a randomly polarized Helium-Neon (HeNe) laser with $\lambda = 632.8$ nm. 
Similarly consistent with expectations, the normalized Stokes vector was found to be $(1.000, 0.998, 0.016, -0.026)$, giving a DOP of 99.9\%. Circular polarization was created by placing a QWP in front of an external cavity diode laser operating at $\lambda = (794.979\pm0.005)~$nm. After recalibration, the average normalized Stokes vector for this state was found to be $(1.000, 0.003, 0.001, -0.986)$ indicating left circular polarized light with a DOP of 98.6\%.

\begin{figure}[ht]
\begin{center}
\includegraphics[width = \columnwidth]{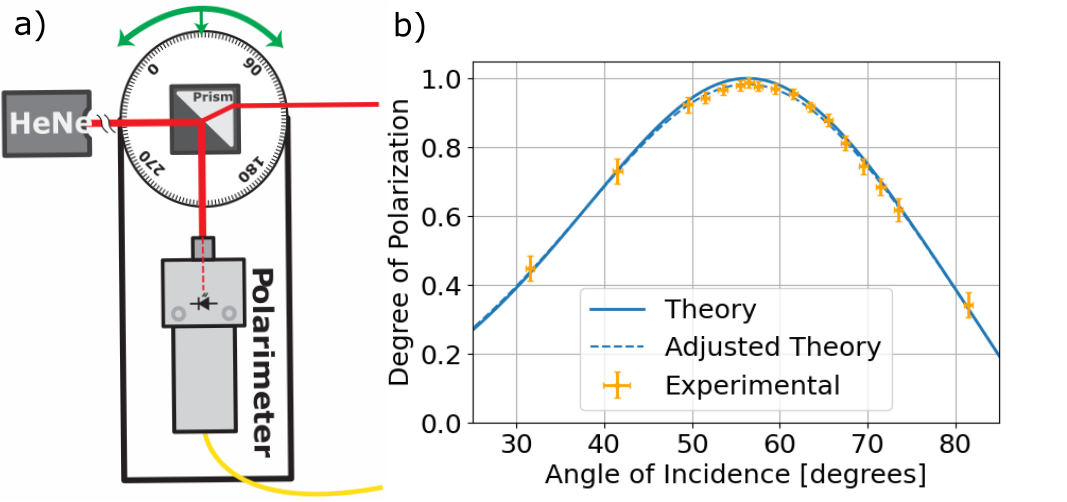}
\end{center}
\vspace{-15pt}
\caption{Demonstration of partial polarization capabilities of the polarimeter. a) Depicts the apparatus used for the polarization by reflection experiment. b) Compares the theoretical and experimentally determined degrees of polarization for light reflected off a prism from a $30^{\circ}$ to a $80^{\circ}$ angle of incidence. The adjusted theory  assumes an angular spread $2\Delta\alpha$ of $5^\circ$.}
\label{fig:polbyref}
\end{figure}

Partially polarized states were created using polarization by reflection. Light from the unpolarized HeNe laser was reflected off a glass ($n \approx$ 1.52) prism at range of angles of incidence (Figure~\ref{fig:polbyref}a). The DOP of the reflected light can be described by: 
\begin{equation}
\label{eq:DOP_polbyref}
P(\alpha) = \left|\frac{R_S(\alpha) - R_P(\alpha)}{R_S(\alpha) + R_P(\alpha)}\right| 
\end{equation}
\noindent where $R_S$ and $R_P$ are the reflectance of S-polarized and P-polarized light obtained from the Fresnel equations~\cite{HechtBook} and $\alpha$ is the angle of incidence with respect to the normal.
Equation 14 is adjusted to account for angular spread by averaging the DOP over a small range about the central measured angle. Figure~\ref{fig:polbyref}b compares the DOP of reflected light measured by the polarimeter to the theoretical DOP and the adjusted DOP for a range of incidence angles. The results show a strong correlation between experiment and prediction; all values are consistent with the adjusted theory. Further discussion of adjustments and assignment of experimental uncertainties can be found in the supplementary material.

\vspace{-10pt}
\subsection{Application: Extremely Affordable Waveplates}
\label{subsec:cheapWPs}

\vspace{-10pt}
A significant barrier to advanced experiments in low budget optics and AMO physics labs is the number of optical components required. Certain experiments such as magneto-optical trapping require many QWPs, each of which can cost hundreds of dollars. It was noted
\cite{savukov2007wave}
that high quality waveplates may be constructed using overhead transparency paper. We investigate this by sending linearly polarized light through transparency paper into our polarimeter and observing the polarization ellipse. Our real time display allows for a quick determination of regions with retardance appropriate for QWPs. 

\vspace{-10pt}
\subsection{Application: Time Dependant Polarization Mapping}
The real-time nature of our SWP allows for \textit{in situ} characterization of time dependent polarization states. The acquisition rate is limited by the mechanical rotation rate of the device, which is typically approximately 100~Hz. This sets the maximum sample rate for non-graphical operation, though a reduced rate is preferable for averaging and numerical accuracy. For real-time graphical display, the update rate is only required to be on the order of Hz, which is easily accomplished in our Raspberry Pi system. 

\begin{figure}[ht]
\begin{center}
\includegraphics[width = \columnwidth]{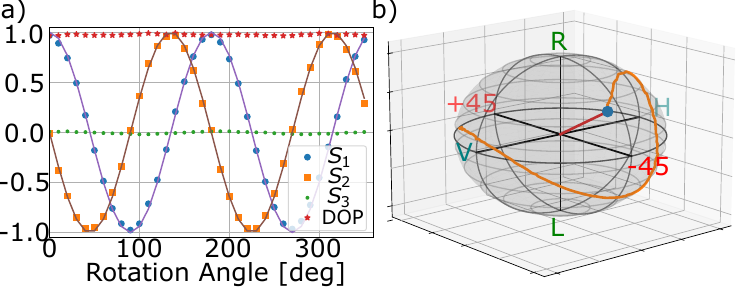}
\caption{Time dependent measurements of the Polarizations state. a) Shows the measured Stokes parameters when an elliptically polarized beam is sent through a rotating polarizer. b) Shows the  trajectory on the Poincar\'{e} sphere of linear polarization through a rotating transparency paper.}
\label{fig:rt_poincare}
\end{center}
\end{figure}

Figure~\ref{fig:rt_poincare}a shows the measured Stokes parameters for regular sample intervals while rotating a linear polarizer with an input circular polarization state from a $\lambda = 650~$nm diode laser. Additionally, the trajectory of a polarization state can be acquired and visualized on the Poincar\'{e} sphere in real-time by keeping a buffer of the last $N$ points and plotting these along with the current location. Figure~\ref{fig:rt_poincare}b shows such an example for acquiring data at 4 fps while rotating a transparency paper by hand. 

\vspace{-30pt}

\section{Conclusions}
We have presented a fully-integrated, affordable polarimeter that can be constructed by anyone with access to basic machining tools. Procedures for calibration and operation were provided to allow for the use of a waveplate with non-ideal retardance, and thus yield a wide range of analysis wavelengths. By extending the analysis to partially polarized light, the device allows for an online visualization of an input optical state of arbitrary polarization. This has potential for visual demonstration in educational settings, and for providing a convenient heads-up display for in-lab calibration procedures. We envision potential applications in determining polarization dependent physical quantities such as concentration of optically active substances in a solution and stress induced birefringence metrology. All supplementary materials for construction and use are provided in a GitHub repository~\cite{SWP_Github}.

\vspace{-10pt}

\section{\label{sec:supplemental}Supplementary Materials}

\vspace{-10pt}
Supporting material for section~\ref{subsec:calib} as well as a bill of materials and construction notes can be found in the Supplemental materials.

\vspace{-15pt}
\begin{acknowledgments}
We thank Chris Secord and Kody Matthews of the University of Victoria, Faculty of Science Machine Shop and Nicolas Braam of the the Dept. of Physics \& Astronomy Electronics Shop for assistance with the design and construction of the device. We acknowledge the support of the Natural Sciences and Engineering Research Council of Canada (NSERC) (grant RGPIN-2021-03289). A. Saji and A. Felipe Ramos were funded through the Mitacs Globalink International internship program.
\end{acknowledgments}
\vspace{-15pt}

\section*{Data Availability Statement}
The data that support the findings of this study are available from the corresponding author upon reasonable request.
\vspace{-10pt}
\section*{References}
\vspace{-10pt}
\bibliography{swp_2024}

\end{document}